Si Nanowire-Array Source Gated Transistors

*Charles Opoku, Radu Sporea, Vlad Stolojan, Ravi Silva, and Maxim Shkunov\**

[\*]     Dr. Maxim Shkunov, Dr Charles Opoku, Dr  Radu Sporea, Dr Vlad Stolojan, Prof Ravi Silva
Advanced Technology Institute, Electronic and Electrical Engineering Department,
University of Surrey , Guildford, GU2 7XH , UK

E-mail:     m.shkunov@surey.ac.uk

Solution processed field-effect transistors based on single crystalline silicon nanowires (Si NWs) with metal Schottky contacts are demonstrated. The semiconducting layer was deposited from a nanowire ink formulation at room temperature. The devices with 230nm thick $SiO_2$ gate insulating layers show excellent output current-voltage characteristics with early saturation voltages under 2 volts, constant saturation current and exceptionally low dependence of saturation voltage with the gate field. Operational principles of these devices are markedly different from traditional ohmic-contact field-effect transistors (FETs), and are explained using the source-gated transistor (SGT) concept in which the semiconductor under the reverse biased Schottky source barrier is depleted leading to low voltage pinch-off and saturation of drain current.  Device parameters including activation energy are extracted at different temperatures and gate voltages to estimate the Schottky barrier height for different electrode materials to establish transistor performance-barrier height relationships.

Numerical simulations are performed using 2D thin-film approximation of the device structures at various Schottky barrier heights. Without any adjustable parameters and only assuming low-p-doping of the transistor channel, the modelled data show exceptionally good correlation with the measured data. From both experimental and simulation results, it is concluded that source-barrier controlled nanowire transistors have excellent potential advantages compared with a standard FET including mitigation of short-channel effects, insensitivity in device operating currents to device channel length variation, higher on/off ratios, higher gain, lower power consumption and higher operational speed for solution processable and printable nanowire electronics.

# 1. Introduction

Solution-based fabrication of field-effect transistors (FETs) using semiconducting 'inks' at low temperatures and on large area substrates is a highly attractive technology for low-cost, lightweight and potentially flexible electronics applications in many areas including: chemical and biological sensors,[1] RFID tags,[2] memory elements,[3] ambient intelligent devices, e-paper and flexible displays.[4]

Semiconducting nanomaterials such as single-crystalline nanowires (NW) have the potential to provide a breakthrough in the area of high performance, low cost device assembly due to high charge carrier mobility comparable to that of their bulk counterparts. The unique aspect ratio allows NWs with a few tens of nanometer diameters to be dispersed in solvents and then processed onto substrates at room temperature to 'bridge' device electrodes. Provided that there is good degree of alignment of these NWs along charge flow direction and high quality ohmic contacts to the semiconducting channel device performance per NW is expected to be comparable to that achieved in traditional single-crystal semiconductor technology.[5] Recent synthetic growth techniques like vapour-liquid-solid[6] (VLS) and supercritical-fluid-liquid-solid (SFLS)[7] methods developed for growing NWs now have the potential to produce industrial quantities of materials. More recently, solution-based nanowire positioning technology has been demonstrated over 16,000 patterned electrode sites on a single substrate.[8] These advances make it feasible for using NWs and active channel elements in low cost scalable assembly techniques.

However, in order to realistically fulfil the potential for semiconducting NW printable electronics, key issues such as contact formation to NW FET channel need to be better understood. Currently, the complex processing steps required to form n (or p) doping to create ohmic contacts in traditional FETs will encounter difficulties in printable NW systems.[9] Attempts to obtain ohmic contacts by growing silicides on Si NW contact areas have improved NW FET performance,[10] though required process temperatures makes this these approaches unsuitable for building NW electronic devices on flexible substrates such as plastics.

A more realistic approach to define contacts in solution processed NW FET structures is the direct deposition of metal contacts onto semiconducting NWs to form source-drain electrodes.[11] Typically these contacts are Schottky[12] with significant charge injection barriers that results from the mismatch of the metal-semiconductor work functions. High Schottky barriers severely limit efficient charge carrier injection into the FET channel and are typically considered to be a major disadvantage in NW FETs.

Despite these limitations Knoch et al.[13] have shown that high Schottky barriers at the contacts in MOSFETs can offer stability in terms of device on-current ($I_{on}$) and immunity from scattering effects in the channel. Using Si NWs as the active channel in Schottky barrier NW FETs, Koo et al.[14] have demonstrated that the small diameter of NWs can offer significantly lower off-current ($I_{off}$) when compared to reference devices with higher channel widths.

In fact Schottky barrier transistors with a particular design of device structures can offer a number of advantages including simple and low temperature fabrication steps, good suppression of short channel effects, as well as eliminating the need for doping and subsequent high temperature activation steps.[13, 14]

In this work, we demonstrate excellent output characteristics for source barrier silicon nanowire-array transistors where the device geometry allows efficient manipulation of depletion area under the source-electrode by the gate field, and describe the physics of device operation when different metals are used to form various source-barrier heights. As device current is mainly determined by the electric field dependence of the source barrier and not by the channel conductance as in typical ohmic contact FETs,[15] abrupt current saturation with drain voltage occurs when the source is depleted of charge carriers and the electric field at the source barrier also saturated.[16] This effect leads to several advantages such as early current saturation of less than 2 volts even at high gate voltages, device current insensitivity to the variation of the channel length and the possibility to overcome short cannel effects that can be exploited in high performance printed electronics based on semiconducting NWs.

## 2. Silicon nanowire based transistors

The Si NWs used in this study were synthesized via the supercritical-fluid-liquid-solid (SFLS) method where toluene was the solvent of choice.[7, 17] The synthesis was carried out at 480 °C and 100 atmospheres in the reaction vessel, ensuring a supercritical phase. A scanning electron microscope (SEM) image of the SFLS grown Si NWs is shown in Figure 1a. The length of the Si NWs typically ranged from 5μm to 40μm. Statistics of nanowire diameter distribution is shown in supporting information. Figure 1b shows a high resolution transmission electron microscope image (HRTEM) of a typical Si NW of 30 nm diameter. The nanowire is a single crystal grown in the [100] direction, as seen from the insert diffractogram, which is typical of a single crystal Si viewed along [111] direction. The HRTEM image also shows a thin amorphous surface layer of about 3 nm thickness.

The as-grown silicon nanowires could be easily dispersed in various organic solvents, and anisole was chosen as a low toxicity solvent that provided good dispersion stability. Nanowires were coated on substrates using various methods including spin-coating, drop-casting, dip-coating and spray-coating. The latter deposition method was preferred due to mainly two factors including compatibility with large surface area coating and the ability to produce partially oriented nanowire layers on substrates.

Photolithographically defined source and drain (*s/d*) electrodes were formed on top of Si NW arrays as described in the experimental section. Figures 1c and d show SEM images of the fabricated NW FETs with nanowires bridging the device electrodes on a Si/SiO$_2$ substrate. The inset (Fig. 1c) is showing the contact metal profile wrapping around NWs. A schematic representation of the NW FET configuration typically of a bottom-gate top-contact is depicted in Figure 1e.

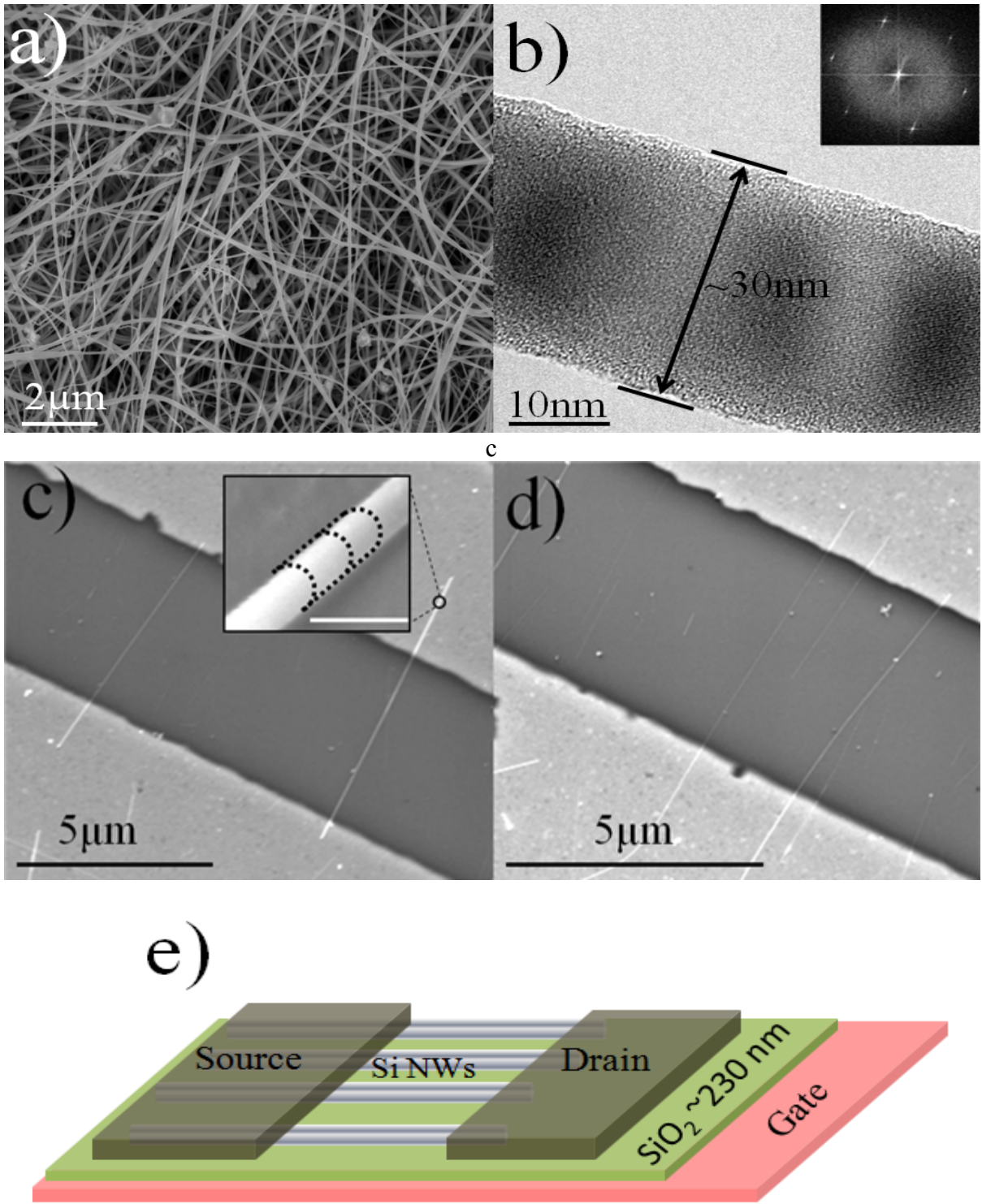

**Figure 1.** a) SEM image of a mesh of Si NWs used in this work. b) HRTEM image illustrating a typical silicon nanowire with ~30nm core diameter. The inset show the reciprocal lattice peaks obtained from Fast Fourier Transforms (FFT) typical of [100] growth direction. c-d) SEM images showing NWs bridging device electrodes on the device substrate, inset in (c) is a SEM image showing metal wrapping over the nanowire, scale bar is ~100nm. e) Schematic representation of a typical transistor structure showing multiple NWs in the channel area.

To assess the impact of Schottky barriers on the Si NW array transistor characteristics, three metals were employed as s/d contacts. Figure 2 shows electrical measurements for devices constructed using gold (Device1), nickel (Device2), and tungsten (Device3) as the *s/d* electrodes. The corresponding work functions of the metals for clean interfaces in a vacuum are 5.1eV (Au), 4.75eV (Ni) and 4.6eV (W).[18]  Due to low processing temperatures, well below the eutectic points for the selected metals, no silicides were formed at the device contacts.

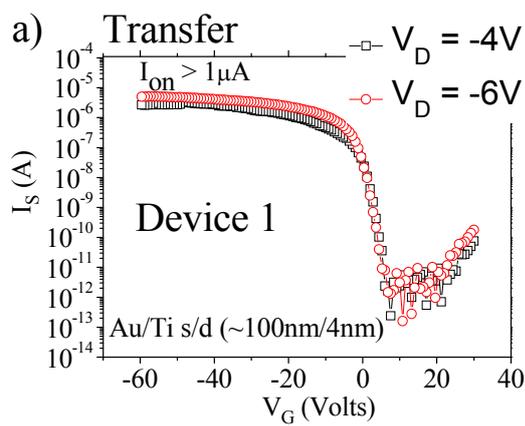
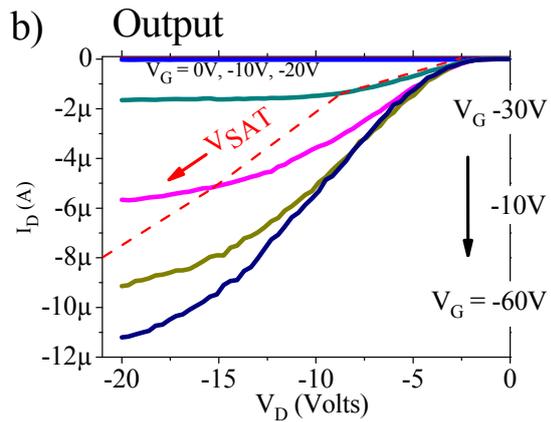
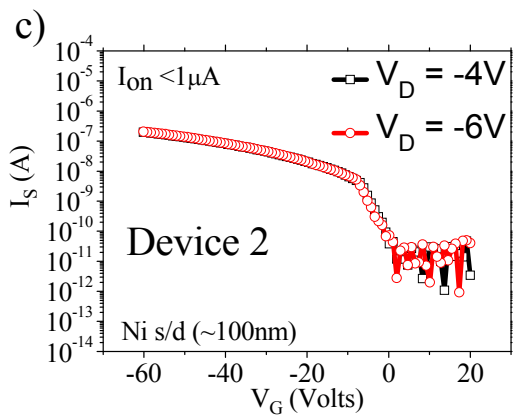
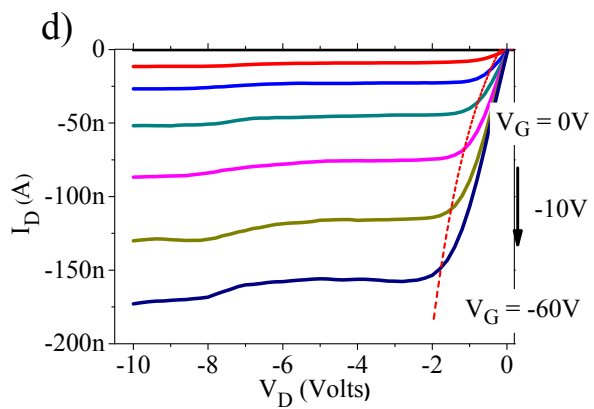

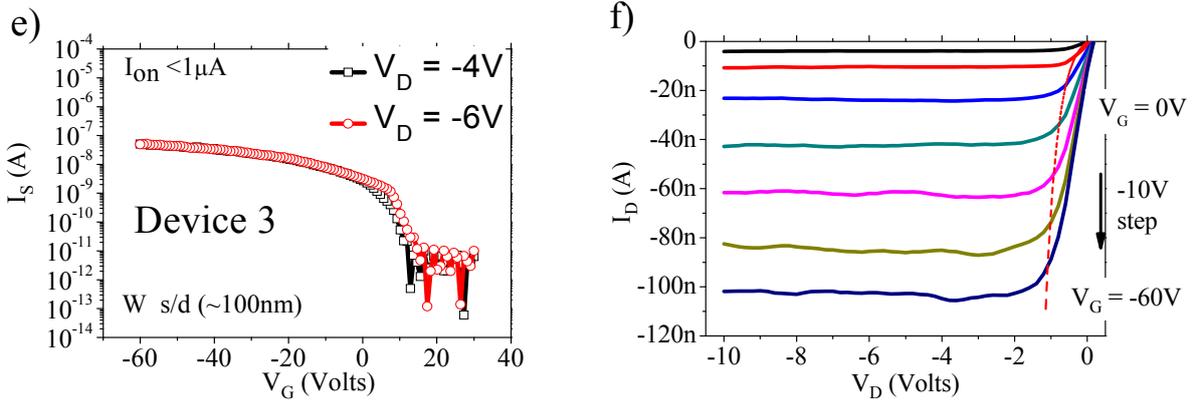

**Figure 2.** Electrical characteristics for three types of Si NW FETs with Au, Ni and W source and drain contacts. Transistor channel length is 5μm. a, c, e) Transfer characteristics ($I_D$-$V_G$) for Devices 1, 2, and 3 respectively. b, d, f) Output characteristics for the same devices. Dashed lines show progression of saturated regime onset.

Assuming no Fermi level pinning and the influence of interfacial states, the upper limits for hole injection barriers ($\Delta\Phi$) can be estimated as the difference between silicon valence band (5.17eV) and metal work functions to be $\Delta\Phi$(Au) ~0.07eV, $\Delta\Phi$(Ni) ~0.42eV, $\Delta\Phi$(W) ~0.57eV.

All Si NW transistors measured in our study demonstrated p-type accumulation behaviour consistent with previous reports.[7] The basic transistor parameters including subthreshold swing (S), peak transconductance ($Gm = dI_D/dV_G$), and *on/off* ratio are extracted and summarized in **Table 1**.

**Table 1.** Transistor parameters for Devices 1-3.

|  | Subthreshold swing $S$ (V/decade) | Transconductance $Gm$ (nS) | *On/off* |
|---|---|---|---|
| **Device1 (Au)** | 1.7 | 150 | $10^7$ |
| **Device2 (Ni)** | 3 | 2 | $2\times10^4$ |
| **Device3 (W)** | 3.5 | 0.4 | $1\times10^4$ |

Transistor characteristics for Devices 2, 3 show striking differences with that from Device 1. Characteristics in Figure 2b resemble FET device output with high contact resistance, indicative by a non-linear current behaviour through the origin for $V_D$ between 0V and 5V. Current saturation voltages ($V_{SAT}$) in this device also follow typical FET pinch-off behaviour for $V_{SAT} \sim V_G\text{-}V_T$. On the contrary, Devices 2 and 3 shows very well defined 'liner' and 'flat saturation' regions with early saturation voltage $V_{SAT}$ in the range of 0.5-2 volts that is very weakly dependant on $V_G$.

Transfer characteristics of Devices 2, 3 (Fig. 2c, and 2e) also show significant differences to Device 1 (Fig. 2a). For example, the on-currents ($I_{on}$) in Device1 (Fig. 2a) at $V_D$ = -4 V and -6V are ~3μA and ~5μA respectively, and clearly show increasing current as $V_D$ is increased. Under the same bias conditions, Devices 2 and 3 (Fig. 2c and 2e) exhibits much lower $I_{on}$ (~0.2μA, and ~0.05 μA). $I_{on}$ for the two devices can also be seen to be insensitive to changes in the drain voltage, and transistors *I-V* curves are practically indistinguishable at $V_D$ = -4V and -6V.

Clearly, operation of Devices 2 and 3 is markedly different from Device 1, and thus require a different treatment compared to the standard field-effect transistor model.[15, 19] Shannon and co-workers have previously investigated a range of amorphous and poly-silicon devices (a-Si and poly-Si) with high Schottky barrier at the source electrode, where the device geometry has been changed to extend the gate electrode under the source.[16, 20] These devices were named source-gated transistors (SGTs) because device current was determined by the effective source barrier height, which was modulated by the electric field induced by the gate.

The present NW transistor structures have a number of similarities with SGTs, in particular, owing to a common silicon gate, there is a significant source-gate overlap, and semiconducting material is effectively 'sandwiched' between the s/d and insulator layer with gate contact. Moreover, substantial work function mismatch between electrode materials (Ni, W) and silicon nanowire valence band is likely to manifest as a high Schottky barrier. To further explain devices operation with Ni and W electrodes, the SGT approach was considered as an appropriate working model.

In a traditional accumulation type FET the gate potential controls the channel conductivity, and consequently device current via the field-effect, and device contacts do not limit charge flow. The FET current saturates when drain region is depleted of charge carrier at the drain electrode at high drain voltages ($V_D > V_G$-$V_T$) and the contact resistance is therefore negligible compared to the transistor channel resistance.[15]

However, in nanowire transistors a perfect ohmic contacts cannot be easily achieved, and as a result, high injection barriers play a major role in device operation.[12]

During a SGT operation, a transistor channel is created by accumulating the majority charge carriers (holes) at the semiconductor-dielectric interface. The device on-current however is no longer controlled by the cannel conductance. On the contrary, current is limited by injection characteristics of the source barrier.[21, 22] When the drain voltage is negative, the source Schottky contact is reverse biased, resulting in very high effective contact resistance that exceeds the resistance of the channel.

The level of current in such SGT devices can be significantly lower than in traditional FETs, depending on the height of the source barrier. When the drain voltage is increased, the space-charge region is increased, and a larger volume of the semiconductor is depleted of holes under the source.[20] At a particular voltage $V_D = V_{SAT}$, the depletion region will reach the insulator interface, and the electric field under the source will stay constant and the transistor current will approach its saturation level (similar to reverse saturated current in a Schottky diode).[15, 23]

Higher transistor current is obtained at higher negative gate voltages due an increase in the electric field at the source and reduction in the effective source barrier height.[21, 22] The new saturated current level is determined in this case by the reverse saturated current of the source Schottky 'diode' with reduced effective barrier height.[16, 20]

## 3. Discussions

### 3.1 Scaling of saturation voltage

The markedly small $V_{SAT}$ for Devices 2, 3 is an unmistakable feature not attributed a normal FET. For a traditional FET, $V_{SAT}$ scales with $V_G$ by the relationship $V_{SAT} \sim V_G - V_T$.[15, 19] Indicated by a dashed line, (Fig. 2b), the saturated voltage point in Device 1 shifts to higher drain voltage values that lay outside the scanned range. On the contrary, output characteristics of Devices 2, 3 show abrupt transition to saturated regime at dramatically lower voltage levels.

To gain qualitative comparison between $V_{SAT}$ behaviour for all three devices, the experimental data were extracted from output characteristics shown in Figures 2 (d,f) for Devices 2, 3 and plotted together with the calculated $V_{SAT}$ values for Device 1 as shown in Figure 3. Since $V_{SAT}$ for the gold-contact nanowire transistor (Fig 2, b) was not reached at $V_G$ -50, -60V, the $V_{SAT}$ vs $V_G$ dependence was evaluated using the FET expression $V_{SAT} = [V_G - V_T]$. As shown in Figure 3, $V_{SAT}$ for Devices 2, and 3 remains small even at higher $V_G$, whereas characteristics expected for Device 1 follow the general FET model.

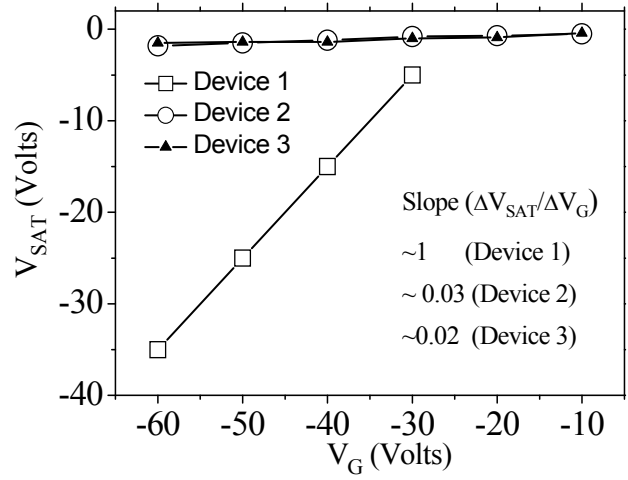

**Figure 3.** Change in $V_{SAT}$ with change in $V_G$ for Devices 1 to 3.

The weak dependence of $V_{SAT}$ has been previously observed for amorphous and poly-silicon source-gated transistors and explained using the so called *dielectric* or *capacitance model*.[16, 21] In the case of a high Schottky barrier (SB) (> 0.3eV) at the source contact in these SGTs, the section of the semiconductor under the source metal is depleted of free charge carriers and behaves as a dielectric. This region of semiconductor and the gate dielectric effectively act as two dielectrics in series, and the gate is coupled capacitively to the depleted source.

In nanowire transistors, contact geometry differs significantly from that of 'planar' this film transistors, and is three-dimensional, and metal layer effectively 'wraps' around the nanowire. A thin depletion layer will be formed at the semiconductor-metal interface for the low work function contacts (nickel and tungsten). We illustrate the formation of an almost fully depleted region under the source electrode in Figure 4(a-c) by showing the cross-section of the nanowire-metal contact at the edge of the source electrode for different drain voltages.

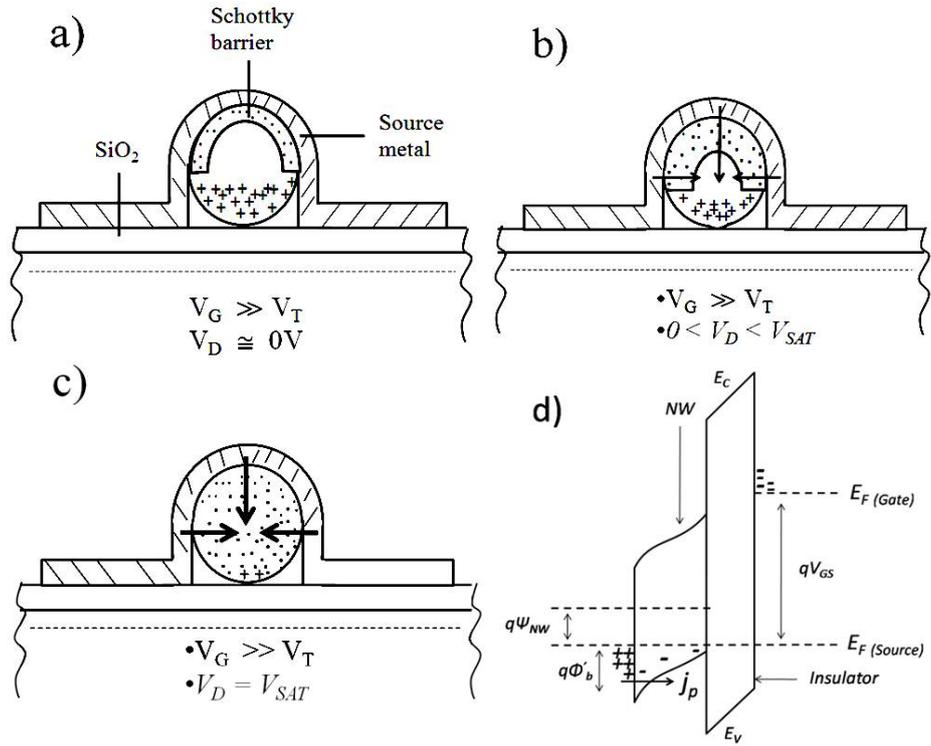

**Figure 4.** a) Depletion profile expected at the source SB in Devices 2 and 3 at $V_D \approx 0V$. b) Depletion region at the metal-NW interface for $0 < V_D < V_{SAT}$. c) Source pinch-off at $V_D = V_{SAT}$. The arrows depict electric field strength at the source. d) Energy band diagram at $V_D = V_{SAT}$.

Initially, for sufficiently large (negative) $V_G$ and small (negative) $V_D$ (Fig. 4b), the NW-insulator ($SiO_2$) interface will be accumulated with holes, and the source Schottky barrier will be under a small reverse bias. Increasing $V_D$ then results in the extension of the depletion region at the source, similar to the operation of the source Schottky diode in reverse bias.

The small diameter of the Si NWs means that the depletion region can extend right from the source metal to the NW-$SiO_2$ interface, provided that sufficient $V_D$ is applied (Fig. 4c). Under these conditions, the source is "pinched off", unlike traditional FETs with ohmic contacts where the drain end pinches off at significantly higher $V_D$ values. The voltage required to establish this source pinch-off is $V_{SAT}$. Further increase in $V_D$ beyond $V_{SAT}$ simply causes the depletion region to expand laterally.[21, 22] Increasing $V_G$ at $V_D = V_{SAT}$ then results in a change of surface potential at the NW surface ($\Psi_{NW}$), causing $\Phi_b$ for holes to drop (Fig. 4d). The change of surface potential ($\Delta\Psi_{NW}$) is thus the same as the change of saturated voltage ($\Delta V_{SAT}$), and can be expressed using Equation 1.[16, 20]

$$\Delta\psi_{NW} = C_i \Delta V_G / (C_i + C_{NW}) = \Delta V_{SAT} \tag{1}$$

Where $C_i$ and $C_{NW}$ represent the capacitance per unit area of the SiO$_2$ gate insulator and the capacitance of the depleted Si NW respectively, and $\Delta V_G$ is the corresponding change of the gate voltage leading to the change in saturated voltage ($\Delta V_{SAT}$).

Having measured $\Delta V_{SAT}/\Delta V_G$ to be around ~0.03 for Device 2, and ~0.02 for Device 3 from Figure 3, the effective thickness ($t_{NW}$) of the depleted section in the NW can be estimated as ~28nm and ~24nm for Devices 2 and 3 respectively. Note that the values of $t_{NW}$ attained are in close agreement with the average diameter of the Si NWs (~30nm) used in this investigation and suggest that the depletion region in the NWs indeed extend from the source metal down to SiO$_2$ interface.

With the calculated $t_{NW}$, the electric field ($E_{NW}$) in the depletion region is simply $V_{SAT}/t_{NW}$.[21] $E_{NW}$ values are estimated to be ~1.8x10$^5$ V/cm and ~7 x 10$^5$ V/cm at $V_G$ = 0V and -60V for Device 2, and ~1.7x10$^5$/cm and ~6x10$^5$/cm for Device 3 respectively. As has already been demonstrated for the amorphous silicon SGTs[22], such high internal fields could mean that nanowire transistors with significant SB at the *s/d* regions will be desirable for high frequency operations.

## 3.2 Gate induced effective barrier lowering

Due to low temperature processing as well as the presence of nanowire surface layers, and even shells, composed of silicon oxide or polyphenyl-silanes,[7] the formation of metallurgical quality clean contacts between the silicon core and metal electrode is very challenging. As a result, the effective barrier height at the semiconductor-metal interface can have various contributions including energetic work function difference, influence of surface states, interfacial layers and others.

Rigorous treatment of nanowire transistor current-voltage characteristics therefore requires knowledge of the exact transport phenomena at the semiconductor-contact interface and also within the transistor channel. Within the SGT model, some simplifications can be made assuming that the transistor current is mostly due to injection over an effective barrier, and channel conductance is always high enough to support this level of current. The effective barrier lowering can occur in nanowire SGT due to a number of reasons including image force potential, surface and interfacial defects, hopping phenomena, field emission, quantum mechanical tunnelling etc.[15] As a result, consideration of an *effective* barrier height and *effective* barrier lowering contact can significantly alleviate the difficulties in the nanowire device treatment.

We start with the simplest expression for current density for a SGT type of device usually given by Equation 2 for a SB in which current transport is dominated by thermionic emission.[15, 21]

$$J = A^* T^2 \exp-\left(\frac{q\Phi_b}{kT}\right) \qquad (2)$$

Where $J = I_S/S$, $S$ is the contact area at metal-semiconductor interface, $A^*$ is the Richardson constant, $T$ is temperature in Kelvin, $\Phi_b$ is the barrier height, $V$ is the voltage bias. The theory predicts that $\Phi_b$ should be voltage independent in the reverse direction.

However, it is well known that the pure thermionic emission model underestimates the reverse bias current[12, 15, 24] given by Equation 2, as it does not account for quantum mechanical tunnelling of charge carriers below the top of the Schottky barrier with increasing reverse bias.

Andrews and Lapselter[25] have shown that Equation 2 is valid if there are no interfacial charges, and evoked an empirical dependence of a barrier lowering in the form $\Delta\Phi_b$ as the missing component to explain their results in metal-silicide Schottky diodes. Shannon[20, 22] have also used a barrier lowering dependence in the form $\alpha E_s$, as a lowering mechanism induced by a large gate field to explain the operational principles in amorphous silicon SGTs. In this case, barrier lowering proportional to the electric field is a good approximation even when there are a number of contributing physical mechanisms.

$$\Delta\Phi_b = \alpha E \qquad (3)$$

In Equation 3, α is an effective barrier lowering constant which for an ideal contact is the effective minimum lateral width of the depletion region permissible for tunnelling by charge carriers. $E$ is the electric field at source barrier. In this regard, the expression for the current in the reverse direction in a general situation involving many contributing mechanisms may be written as:[22]

$$J \approx K \exp -q\left(\frac{\Phi_b - \alpha E}{kT}\right) \qquad (4)$$

Where *K* is a constant: so for a given value of *E* the current increases by *exp(αE/kT)*. Whilst this soft reverse bias effect is an undesirable feature in Schottky diodes, it is the process which describes current modulation in a SGT.

To determine the barrier height as well as the change in the barrier height with the gate voltage $V_G$, variable temperature I-V measurements were conducted on all three devices. Figure 6 shows Arrhenius plots (in the form $I = I_o e^{-(E_a/KT)}$)[21] extracted from these measurements and presented in the form $\ln I_S$ vs. $1/kT$ at $V_D = -6V$. The slopes of the plots yield the activation energies ($E_a$) for several gate voltages.

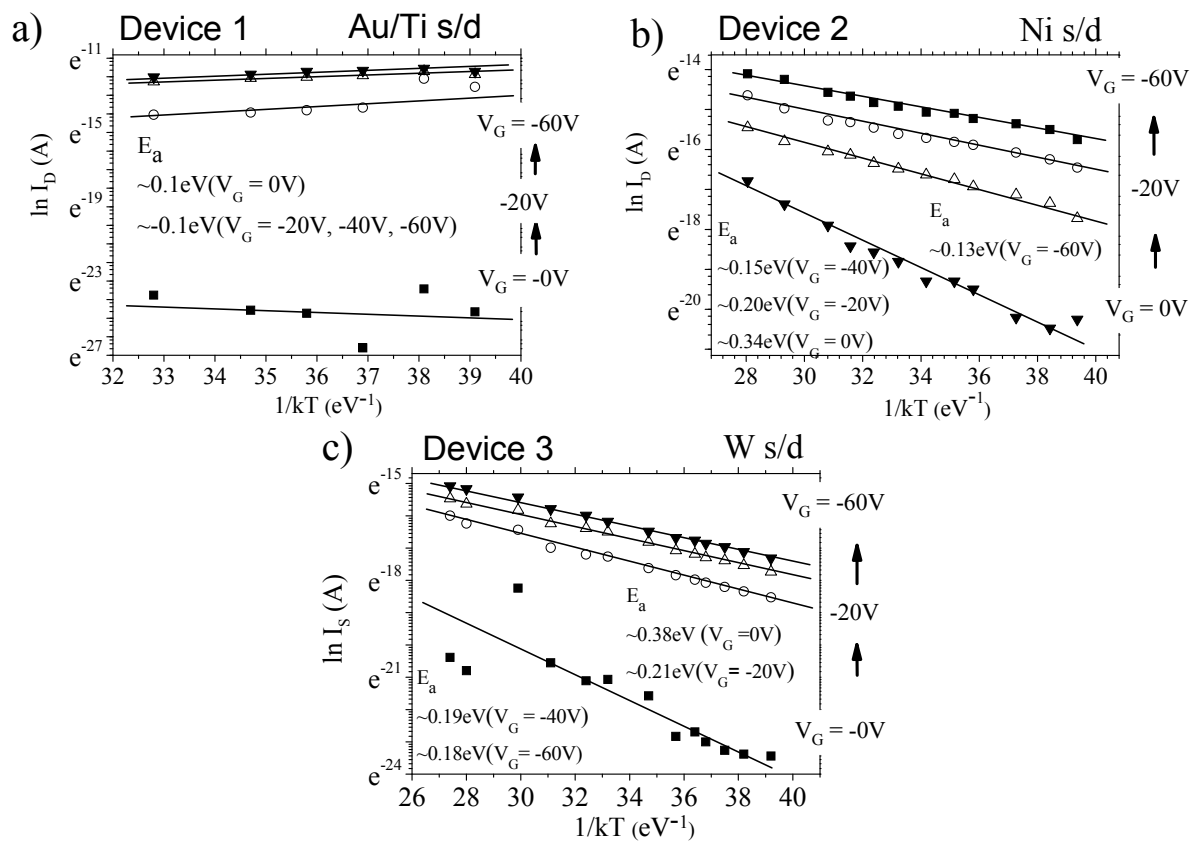

**Figure 6.** Arrhenius plots for Devices 1, 2, and 3 at $V_G$ from 0V to -60V in steps of -20V. a) Device 1(with Au s/d contact). b) Plots for Device 2 (Ni s/d contacts). c) Plots for Device 3 (with W as s/d contacts).

The $E_a$ values extracted at $V_G = 0V$ are ~0.1eV, 0.34eV, and 0.38eV for Devices 1, 2 and 3 respectively. The data presented for Device 1 in Figure 6a confirms a comparatively low Schottky barrier height and explains the more normal FET behaviour of the device exhibiting high drain currents. For this device, $E_a$ is shown to rapidly fall from a positive value of +0.1eV (at $V_G = 0V$), to a negative value of about -0.1eV at all negative gate voltages ( -20V to -60V). This behaviour is typical of standard FETs where carrier scattering at high temperatures reduces carrier mobility.[19] In essence, the negative $E_a$ at higher $V_G$ indicates that the current through this device is determined by the channel conductivity.

$E_a$ values of 0.34eV and 0.38eV for Devices 2, and 3 (Fig. 6b and 6c respectively) at $V_G$=0V are lower than expected for a lightly doped p-type Si with Ni-Si and W-Si as Schottky contacts, but the presence of native oxide on the NW surface may be the reason for this variation.[15, 25, 26] Crucially, $E_a$ is clearly seen to fall with increasing $V_G$ by ~3meV per gate volt to a minimum value ~ 0.13eV and ~0.18eV (at $V_G$ = -60V), for Devices 2 and 3. The reducing $E_a$ with $V_G$ confirms that the source SB is being lowered by the gate field.

Assuming that there is a uniform emission of charge carriers over the reversed source barrier, an effective barrier lowering constant ($\alpha$) can be obtained using Equation 4. From this equation ($\alpha$) is calculated to be ~1.4nm and ~0.8nm for Devices 2 and 3 respectively. Typical values for $\alpha$ in n-type bulk silicon are $\approx$ 2-5nm,[15] but in this case because hole have larger effective mass compared with electrons we expect lower values which are indeed measured. Conversely, interfacial layers will also lower the measured values of $\alpha$.[27]

## 4. Modelling current transport in Si NW transistors

In support of measurements, 2-D numerical simulations have been performed using Silvaco Atlas on a thin-film approximation of the structure. The Si NWs in the channel are collectively described as a 25nm-thick thin-film of lightly p-type silicon with low defect density. A $SiO_2$ layer of 230nm separates the semiconductor from the gate electrode, and the channel length ($L$) is set to 2.5μm. A SB model is used to describe the source contact with the height of the source barrier being set at: 0.37eV. Output and transfer characteristics computed using this 2-D structure is shown in Figure 7. It is seen that the drain current is modulated by the gate voltage (between -20V and -60V). This is due to the gate field acting on the source barrier, and effectively lowering it.

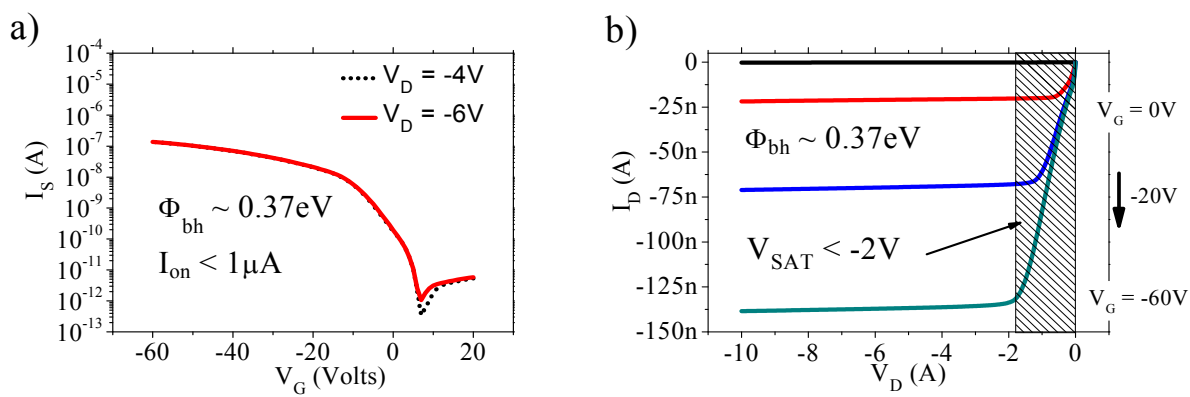

**Figure 7.** Simulated transfer (left) and output (right) characteristics for a thin-film structure approximating the SB Si NW FETs. The Schottky barrier height at the source was set at 0.37eV for holes.

The output plot also reveals the small change in $V_{SAT}$ with $V_G$ and the small drain-field dependence on $I_{SAT}$ (low output conductance). These features show good correlation with the experimental work presented for the Si NW FETs shown in Figure 2. The small change of $V_{SAT}$ with $V_G$ allows the devices to dissipate less power than a conventional FET operating at the same currents for improved energy efficiency.[22] The latter is an essential performance factor in analogue amplifiers and signal processing stages and could enable high performance analogue blocks based on NWs to be integrated in circuits made with low cost technologies.[28]

Numerical analysis of carrier concentrations at the source and in the channel reveals the saturation mechanisms in the simulated structures (Fig. 8). An applied drain voltage creates a depletion region at the edge of the reverse-biased source barrier. As the drain voltage reaches $V_{SAT} = V_G (C_i/C_i+C_{NW})$, where $C_i$ and $C_{NW}$ are the respective capacitances per unit area of the insulator and nanowire, the semiconductor pinches off under the source and the current saturates.

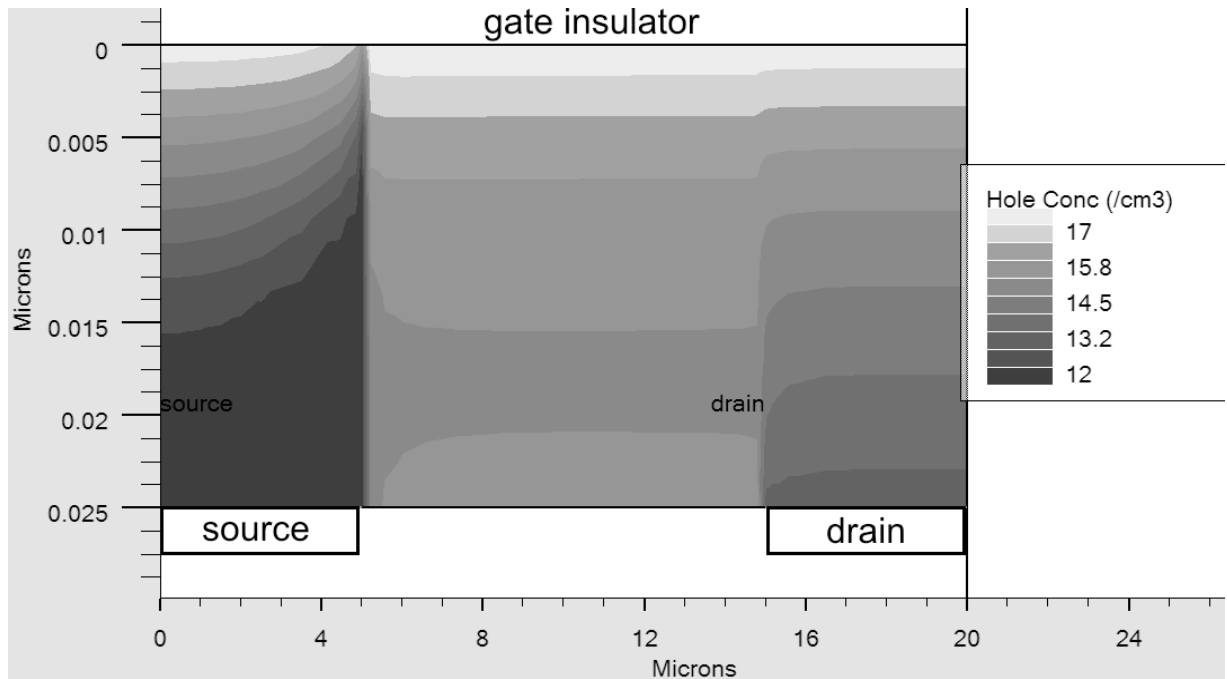

**Figure 8.** Numerical simulation of the structure showing the concentration of majority carriers (holes) in the semiconductor for $V_{SAT} < V_D < V_{SAT\_FET}$ and the pinch-off at the source.

In a conventional NW FET, drain current saturation would occur at $V_{SAT\_FET} = V_G - V_T$. Because a thick insulator and a relatively small diameter nanowire are used, the magnitude of $V_G$ is large but $V_{SAT}$ is much smaller than $V_{SAT\_FET}$. When $V_D$ is between $V_{SAT}$ and $V_{SAT\_FET}$, the semiconductor region between the source and the drain behaves as a FET channel in the linear region ($V_D<V_{SAT\_FET}$) but the current saturates at low $V_D$ due to the source pinch-off. Changes of $V_D$ in this regime have little impact on the magnitude of the drain current, as the pinched-off source region screens the source barrier from the penetration of the electric field generated by the drain voltage.

Additional simulations with barriers of 0.19eV and 0.72eV have been performed and the barrier lowering caused by the gate field for the simulated structures is illustrated in Figure 9.

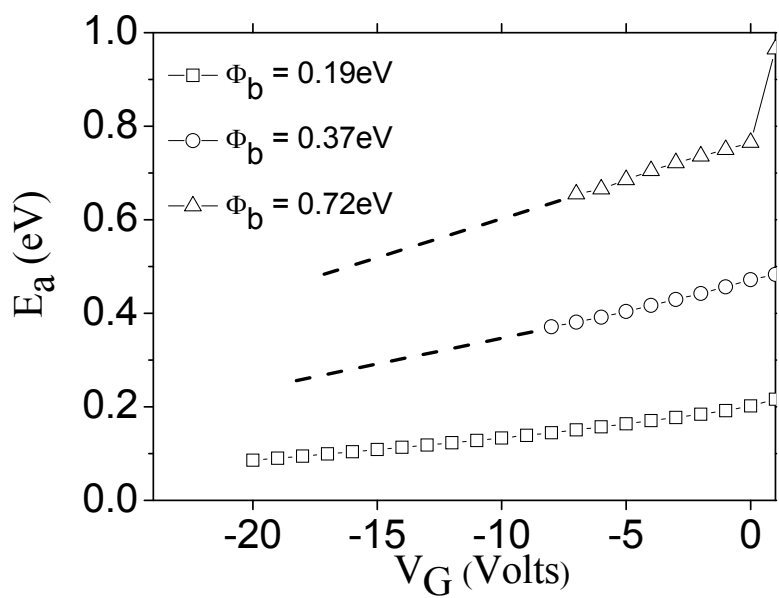

**Figure 9.** Activation energy of the drain current vs. applied gate bias for the simulated structures at three source-barrier heights. The effective barrier height decreases with applied gate field.

Increasing the gate field causes a reduction in $E_a$ for current transport across the source SB. This correlates well with measured data (Fig. 6). In the extreme case of a low SB being pulled down by the gate field, the activation energy can become negative (Fig. 6a), hinting at the fact that the barrier has stopped being a controlling mechanism for the current and the structure operates as a normal FET.[15] The numerical simulation of a planar structure represents only a first-order approximation of the fabricated device geometries. However, it provides a model of the physical processes involved, thus leading to good agreement in the first order based on the experimental data.

5. Conclusions

We have demonstrated silicon nanowire array field effect transistors where the active layer was processed from solvent based 'ink'. The nanowire deposition was conducted at room temperature and the process is compatible with large-area fabrication techniques envisioned for low-cost printed electronics.  Device electrodes were deposited in a more traditional way using photolithography. Electrical and simulation analysis based on three silicon nanowire-array field-effect transistors with Au, Ni or W as source and drain contacts have been made. With Au contacts, providing low injection barrier, the characteristics showed standard field-effect transistor behaviour with operating currents exceeding 1μA. Obtained activation energy for holes (~0.1eV at zero gate voltage) was consistent with expected Schottky barrier height.  In contrast, devices with Ni or W as source and drain contacts, collectively exhibited lower operating currents resulting from the formation of a source barrier on the nanowire. Barrier heights were 0.34eV and 0.38eV for the Ni and W respectively. These Schottky source barriers meant that such devices did not conform to conventional transistor rules, and were characterised by two key defining features: insensitivity of the on-current to the magnitude of the drain bias above saturation, and very low saturation voltages even at high gate voltages. Further investigation of the Ni and the W devices revealed entirely different operating mechanism that could be explained using the source-gated transistor concept in which saturation occurs when the reverse biased source barrier depletes the semiconductor below it. Measurements of the activation energies for current transport showed that the effective barrier height of the source is pulled down by the gate which supports the gate induced barrier lowering mechanism.

The change in saturation voltage with the gate voltage for transistor made in this way can be some 50 times less than in conventional FET transistors. The small change in saturation voltage with the gate will allow silicon nanowire field-effect transistors with source barriers to dissipate less power than a normal field-effect transistor operating at the same currents. These features suggest that solution based assembly of field-effect transistors using nanowire-arrays at low temperature can offer improved efficiency.

Using the capacitance model originally proposed by Shannon and Gerstner[20] in source-gated thin-film transistors, the depletion region at the reversed source in the Ni and W Schottky barrier devices is shown to be on the order of the average nanowire diameter, and the gate coupling effect is established near the source electrode during transistor operation. Changing the gate voltage then changes the electric field at the source barrier interface and lowers the effective height of the Schottky source barrier. The lowering of the barrier was verified by looking at the activation energies when a large negative gate voltage was applied. It was found that a reduction of as much as 60% could be achieved.

Furthermore 2D simulation have successfully reproduced experimentally observed features of the source-gated transistors including early saturation, 'flat' transistor current-voltage curves after pinch-off, subthreshold behaviour, and even the magnitude of the current at similar biasing, assuming a comparable barrier height (0.37eV).

These properties are ideally suited for printed electronic approaches in source-gated transistor fabrication. Device printing and deposition steps as currently being demonstrated for organic-based transistors,[29, 30] offer only limited reproducibility for printed electrode line widths, and also dielectric thickness variations. Since SGT current is mostly dependant on the source electrode barrier height, and not on the channel conductance, these devices should be practically insensitive to a range of channel length variations, as long as effective contact resistance far exceeds the channel resistance. Moreover, the source-gated transistor small sensitivity of the gate dielectric thickness variations can mitigate the problem of insulator thickness uniformity.

To realise full potential of the source-gated transistor design, the sub micron transistor channel lengths needed can be achieved by printing, as demonstrated by ink-jet printing of two inks with self-dewetting properties.[31] With sub-micron channel length, source-gating technique can give a double advantage, firstly the insensitivity of the SGTs to short channel effects, and secondly, the high internal field achieved make these devices ideal in high frequency operations when compared to traditional field-effect transistors.

## 6. Experimental

*Materials:* The Si NWs used in this study were synthesized via the supercritical-fluid-liquid-solid method according to published procedures, and were not intentionally doped.[7, 17] For material characterisation, the reciprocal lattice peaks obtained from fast Fourier transforms (FFT) of a high resolution TEM (HRTEM) image was used to confirms that the Si NWs are single crystalline with predominant [100] growth direction. The average NW diameter was found to be ~30nm. Typical nanowire diameter distribution, as determined from TEM images, is shown in supporting information. From Scanning electron microscope images, typical length of the Si NWs used in this work ranged from a few micrometers to tens of micrometers long.

*Device fabrication*: To fabricate devices, Si NWs were initially suspended in anisole and then were solution cast onto clean $Si/SiO_2$ substrates. The $SiO_2$ was ~230nm thick and was thermally grown on top of a heavily doped ($n^{++}$) Si. The substrate oxide layer served as the gate insulator and Si as the common gate electrode respectively. Photoresist layers were coated directly on top of nanowire-covered substrates. Photolithography (lift-off) was used to open windows in a photoresist layer. Prior to depositing the source-drain (s/d) contacts, substrates were immersed in diluted HF (~8% HF in DI-water) for up to 10 seconds to remove native oxides, and then rinsed in DI-water and dried under a flow of $N_2$ gas. Metals were sputtered over substrates and lift-off in acetone was performed to complete FET structures. Finally FETs were annealed at ~250°C in $N_2$ atmosphere for 10 minutes to improve the metal/nanowire contacts, and to remove residual solvents. SEM images were used to confirm NW density of ~100 NW/mm$^2$ in the channel of completed device. Typically

NW transistors contained approximately 4 nanowires between interdigitated source-drain electrodes for devices with 800 μm channel width.

*Device characterisation*: I-V characteristics were obtained using Agilent 4142B Modular DC Source/Monitor system. Variable temperature measurements have been performed on a Linkam LTS350 stage equipped with a temperature controller. All measurements have been carried out in a $N_2$ filled glovebox to minimise influence of moisture and other species present in air on experimental data.

**Acknowledgements**

CO thanks EPSRC UK for the provided support ( CASE/CNA/07/79) and MS acknowledges support from EPSRC UK grant (EP/I017569/1).

**Supporting information:**

Differences between field-effect transistors and source-gated transistors are summarised in Table S1 below.

| Field-effect transistors | Source-gated transistors |
|---|---|
| $I_{SAT} = \dfrac{\mu W C_i}{L}(V_G - V_T)^2$ | $I_{SAT} = K \exp \dfrac{-q(\phi_B - \alpha E_s)}{kT}$ |
| $V_{SAT}$ scales with $V_G$ | $V_{SAT}$ shows weak dependence on $V_G$ |
| $V_{SAT} = (V_G - V_T)$ | $V_{SAT} = \dfrac{C_{insulator}}{(C_{insulator} + C_{NW})}(V_G - V_T) + K$ |
| Suffer from short channel effects | Excellent for short channel |
| Require thinner dielectrics for short channel devices (maintain constant field) | Can work with short channels and thick insulators |

Table S1. Comparison of operational principles for traditional FETs and SGTs. Notations are explained in the main text.

As–made (supercritical-fluid-liquid solid synthesis ) nanowires are characterised by distribution of lengths and diameters. Figure S1 shows statistical variation of silicon nanowire *diameters* based on TEM measurement results for 100 nanowires.

Length distribution of nanowires (5μm to 40μm ) was determined from both optical microscope and SEM measurements.

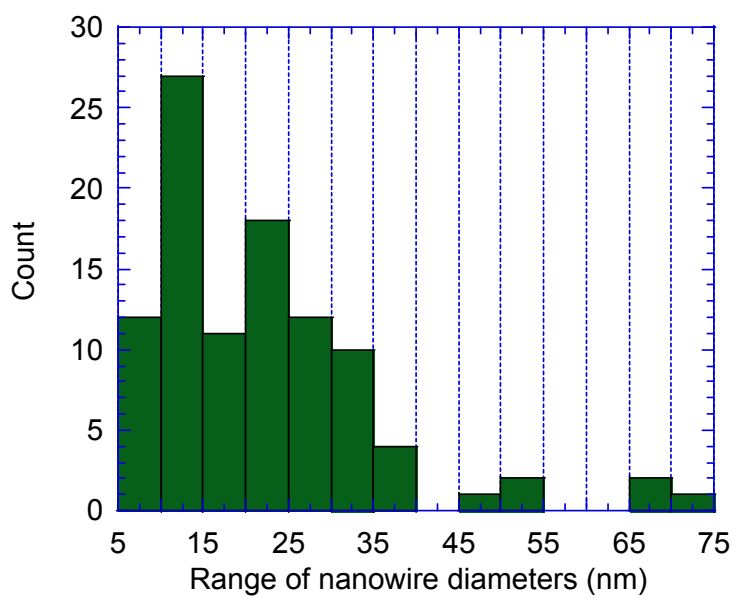

Fig. S1  Distribution of nanowire diameters showing how often particular range of diameters is appearing in a typical TEM image.

Activation energy dependence on the magnitude of the electric filed for devices with high Schottky barriers with nickel and tungsten electrodes. Data was extracted from variable temperature I-V measurements conducted on all three devices.

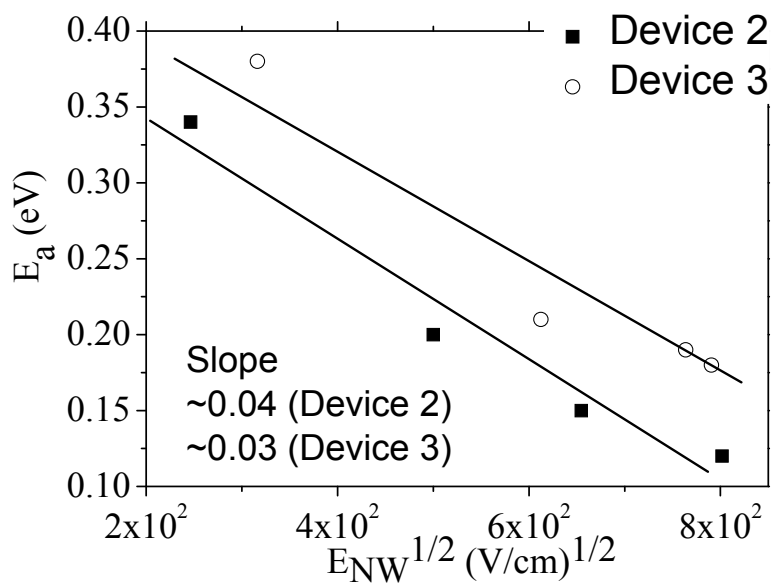

Fig. S2  Activation energy ($E_a$) vs. square root of the electric field ($\sqrt{E}$) measured for Devices 2, and 3. The electric field has been calculated by $E = V_{SAT}/t_{NW}$. $V_{SAT}$ is the saturation voltage at a particular gate bias. $t_{NW}$ is the effective thickness of the depletion region in the NWs, measured using the capacitance model. $\Delta E_a / \Delta \sqrt{E}$ is measured as 0.04 and 0.03 respectively.

2-D numerical simulations results (Silvaco Atlas) were performed for a thin-film approximation of the nanowire structures as described in that main text. The Si NWs in the channel are collectively described as a 25nm-thick thin-film silicon. Simulated characteristics of SGTs for 0.19eV and 0.72ev barrier heights are shown below (Fig. S3)

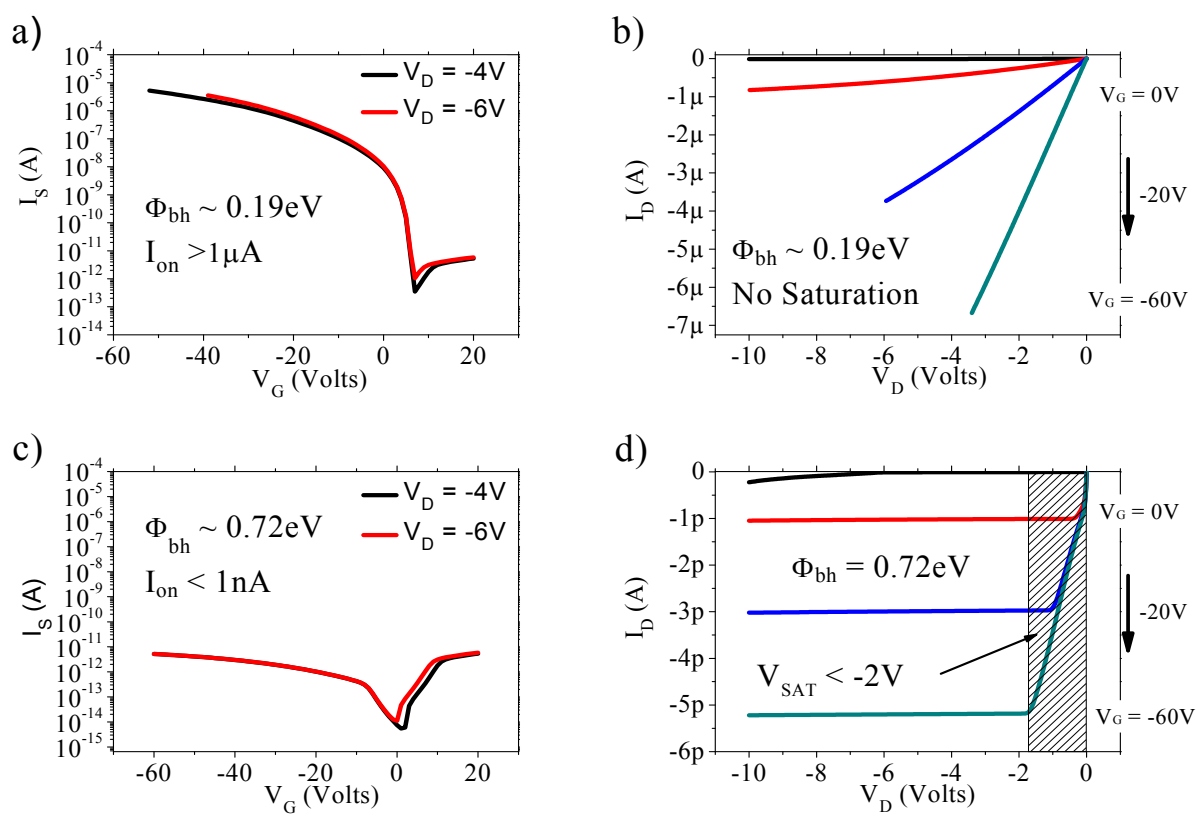

Fig. S3 Simulated transfer (left) and output (right) characteristics for a thin-film structure approximating the Schottky barrier Si NW transistors. The SB height (for holes) at the source electrode was set at: a,b) 0.19eV; c,d) 0.72eV. The device characteristics in (a) and (b) exhibit higher currents and do not saturate. This is explained by the interplay between the low potential barrier for holes and the lowering effect which the gate field has on the source barrier. Due to barrier lowering, at high $V_G$, $\Phi_b$ all but disappears and the structure behaves a normal FET, since the semiconductor cannot be depleted under the source in the absence of a reverse-biased source barrier. In the process, the output conductance at low $V_D$ is compromised. Nevertheless, the presence a potential barrier at the source is relevant, as the off-current ($I_{off}$) is low.
The structure with high source barrier for holes (c) and (d) has a predictably low current but a very sharp transition into saturation and very low output conductance in saturation.
(References are given for the main text)

Simulated cross-section of SGTs showing charge carrier density for two different biasing conditions

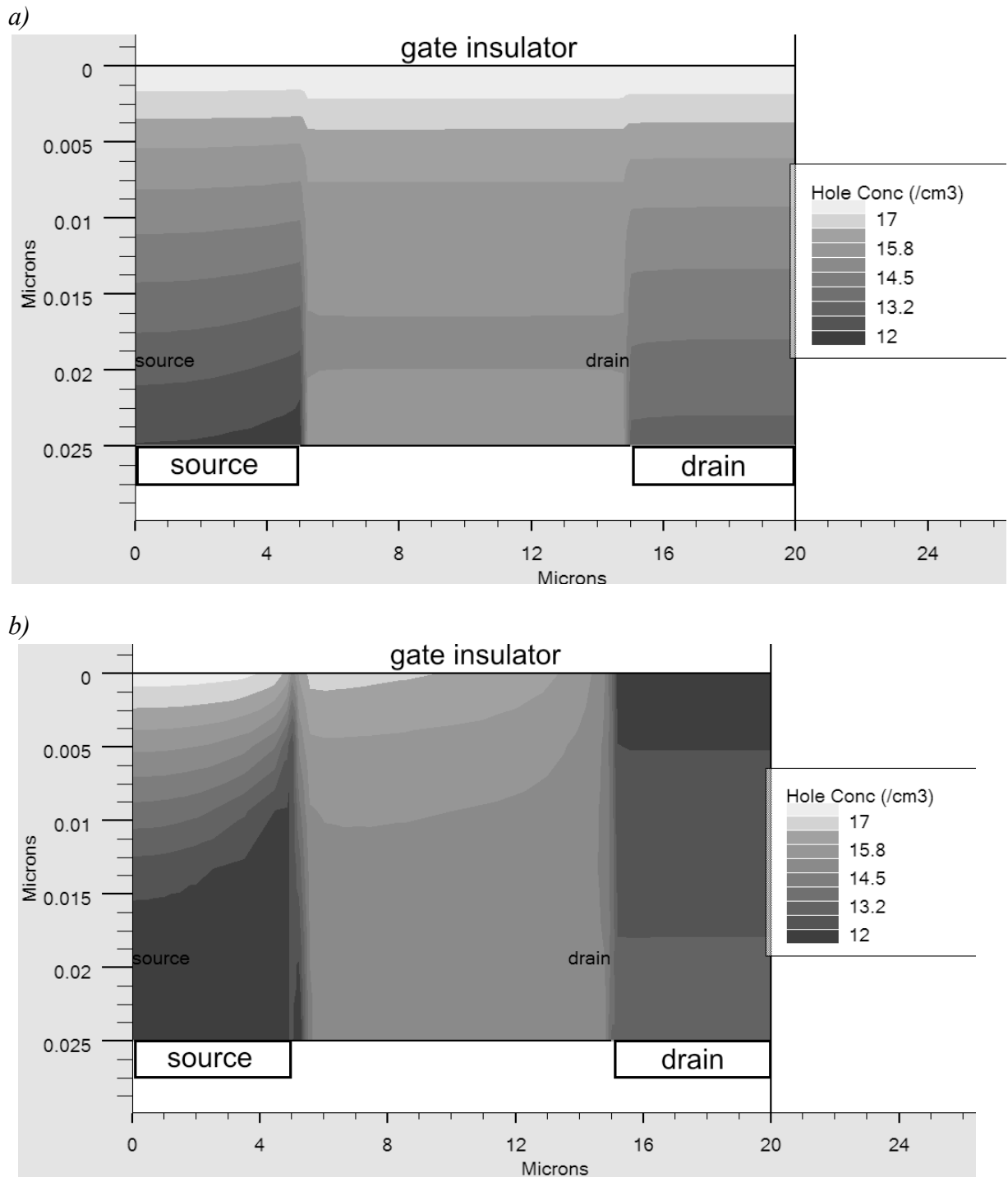

Fig. S4 Numerical simulations of the structures showing the concentration of majority carriers (holes) in the semiconductor for different drain biasing conditions, at the same gate voltage: a) low $V_D$; b) $V_D > V_{SAT\_FET} = V_G - V_T$. In (a), the charge in the source region begins to be depleted by the drain field, but the semiconductor-insulator interface is still in strong accumulation over the whole length of the source region and the device is in the linear

regime. Plot (b) shows the channel of the FET which forms between the source and the drain is pinched off at the drain. Saturation occurs at both the source and the drain when $V_D = V_{SAT\_FET}$. This leads to a very flat saturation characteristic, as subsequent increases in $V_D$ are dropped at the drain end of the channel and do not affect the potential barrier at the source.